\documentstyle[twocolumn,prl,aps]{revtex}
\begin{document}
\draft
\title{Bose-Einstein Condensate: a Superposition of Macroscopically 
Squeezed States}
\author{Patrick Navez\thanks{Also: 
Institute of Materials Science, Demokritos N.C.S.R.,
POB 60228, 15310 Athens, Greece}\thanks{
Email: Patrick.Navez@quantum.physik.uni-potsdam.de}}
\address{Institut f\"ur Physik, Universit\"at Potsdam,
Am Neuen Palais 10, D-14469 Potsdam, Germany}
\date{\today}
\maketitle
\begin{abstract}
We study the ground state of a uniform Bose gas at zero temperature in the
Hartree-Fock-Bogoliubov (HFB) approximation. We find a solution of the 
HFB equations which obeys 
the Hugenholtz-Pines theorem. This solution imposes 
a macroscopic squeezing to the condensed state and as a consequence 
displays large particle number fluctuations. Particle 
number conservation is restored by building the appropriate U(1) invariant 
ground state via the superposition of the squeezed states.
The condensed particle number distribution of this new ground state is calculated 
as well as its fluctuations which present a normal behavior.
\end{abstract}
Since the recent observation of Bose-Einstein condensation in dilute gas 
of trapped atoms
\cite{experiments}, the theoretical developments of dilute Bose gas have 
regained a considerable interest. Among all these developments, the 
mean field Hartree-Fock-Bogoliubov (HFB) approach provides a framework 
which allows to describe the properties of a dilute condensate in 
any regime of temperature \cite{Hohenberg,Griffin,Burnett,Siggia}. 
But, according to 
previous works \cite{Hohenberg,Griffin}, the excitation spectrum 
in the approximate HFB approach is not gapless 
in the limit of a large particle Bose gas since it does not fulfill the 
Hugenholtz-Pines (HP) theorem. At least one supplementary approximation 
should be made in order to satisfy this theorem. Among those there are 
the "Bogoliubov" approximation \cite{Huang,Fetter} valid at zero 
temperature and the "Popov" approximation valid at finite temperature 
\cite{Popov}. On the other hand, both these approximations have the 
disadvantage that, contrary to the HFB one, 
they cannot be deduced from a variational principle.
As a consequence, there is no stationary ground state in both 
these approximations but a ground state which exhibits 
a quantum phase diffusion \cite{Lewenstein}. 

In this Letter, we propose to reconsider the variational HFB approach by 
investigating another kind of ground state solution both satisfying 
the HFB equations and the HP theorem. 
Solution which differs from those discussed previously \cite{Hohenberg,Griffin} 
can indeed be found if we 
admit that the mean value of the two atoms 
amplitude $\langle\hat \psi_0 \hat \psi_0\rangle$ of the condensed part field 
operator $\hat \psi_0$ is macroscopically different from the mean value 
of the atoms number $\langle\hat \psi_0^\dagger \hat \psi_0\rangle$.  
This is possible if the condensed part state is macroscopically 
squeezed in the same way as we squeezed photons states in 
quantum optics \cite{Walls}. 

Another complication with the HFB approach is the assumption that the 
$U(1)$ symmetry of the ground state is broken 
\cite{Gardiner}. Thus, we have to face the problem of dealing with states 
of indefinite particle number but instead having a well defined phase. 
These states are certainly not physical since in any experiment the Bose gas 
should conserve its total particle number. This problem is important since  
the macroscopically squeezed ground state we propose suffers 
from prohibitively large 
particle number fluctuations. We solve the inconvenience by taking 
as a better ground state an adequate 
superposition of squeezed states with a fixed particle number 
\cite{Huang}.

The effective many body Hamiltonian of the uniform Bose gas is given by
($\hbar=1$):
\begin{eqnarray}\label{H}
\hat H=
\sum_{\vec{k}} {\vec{k}^2 \over 2m}
\hat{c}^\dagger_{\vec{k}} \hat{c}_{\vec{k}}
+{1 \over 2V} \sum_{\vec{k},\vec{k}',\vec{q}}
{\cal V}_{\vec{q}}  \hat{c}^{\dagger}_{\vec{k}+\vec{q}}
\hat{c}^{\dagger}_{\vec{k}'-\vec{q}}
\hat{c}_{\vec{k}}\hat{c}_{\vec{k}'}
\end{eqnarray}
where $\hat{c}_{\vec{k}} (\hat{c}^\dagger_{\vec{k}})$ is the annihilation 
(creation) operator expressed in the $\vec{k}$ momentum space with 
periodic boundary conditions
($\hat \psi (\vec{r})= 
\sum_{\vec{k}} \exp(i\vec{k}.\vec{r}) \hat{c}_{\vec{k}}/\sqrt{V})$, 
$m$ is the atomic mass, ${\cal V}_{\vec{q}}=\int d\vec{r}\, {\cal V}(\vec{r})
\exp(-i\vec{q}.\vec{r})$ are the Fourier components of the effective
interaction 
pseudopotential ${\cal V}(\vec{r})=(4\pi a/ m) \delta^{3}(\vec{r})
(\partial/\partial r)r$,
and $a$ is the scattering length \cite{Huang,Fetter}.

The HFB ansatz for a homogeneous gas is \cite{Siggia}:
\begin{eqnarray}
|\theta\rangle_g&=&
\hat D^{\dagger} \hat S^{\dagger} |0\rangle
\\
\hat D^{\dagger}&=&\exp[N_c^{1/2} (e^{i\theta} 
\hat{c}^{\dagger}_{\vec{0}}-e^{-i\theta} \hat{c}_{\vec{0}})]
\\
\hat S^\dagger &=&\exp[\sum_{\vec{k}} s_{\vec{k}}
(e^{2i\theta}\hat{c}^{\dagger}_{\vec{k}}\hat{c}^{\dagger}_{-\vec{k}}
-e^{-2i\theta}\hat{c}_{\vec{k}}\hat{c}_{-\vec{k}})/2]
\end{eqnarray}
$\hat D^\dagger$ 
is the unitary displacement operator generating a coherent state in the 
zero momentum mode with a mean number $N_c$. It has the property:
\begin{equation}\label{D}
\hat D \hat{c}_{\vec{0}} \hat D^{\dagger}=
\hat{c}_{\vec{0}}+e^{i\theta}N_c^{1/2}
\end{equation}
$\hat S ^\dagger$ is the unitary
squeezing operator and creates particles in pairs 
$\hat{c}_{\vec{k}}^\dagger\hat{c}_{-\vec{k}}^\dagger$ \cite{Huang,Walls}.
Parameters controlling the population of pairs are 
denoted $s_{\vec{k}}$ while 
$\theta$ is an arbitrary phase factor.
This operator has the Bogoliubov transformation property:
\begin{equation}\label{S}
\hat S \hat{c}_{\vec{k}} \hat S^{\dagger}=
\cosh(s_{\vec{k}}) \hat{c}_{\vec{k}} +
\sinh(s_{\vec{k}}) e^{2i\theta}\hat{c}^{\dagger}_{-\vec{k}}
\end{equation}
The mean particle number is given by the average value 
$\langle \hat N \rangle$ of 
$\hat N= \sum_{\vec{k}} \hat{c}^\dagger_{\vec{k}} \hat{c}_{\vec{k}}$
where we denote
$\langle \hat A \rangle=_{\ g}\!\! \langle\theta| \hat A |\theta\rangle_g$. 
Using the transformations properties (\ref{D}) and (\ref{S}) and the unitary 
properties $\hat D \hat D^\dagger =\hat 1$ and 
$\hat S \hat S^\dagger =\hat 1$,
we get after a normal ordering of the operators 
$N=\langle \hat N \rangle = \sum_{\vec{k}}N_{\vec{k}}$,
$N_{\vec{k}}=\delta_{\vec{k},\vec{0}}N_c+(\cosh(2s_{\vec{k}})-1)/2$. 
Similarly, we get the following expression for the ground state energy
\cite{Siggia}:
\begin{eqnarray}
E_g=\langle\hat H\rangle=
\sum_{\vec{k}}{\vec{k}^2 \over 2 m}N_{\vec{k}}
+{2\pi a \over mV}(N^2-2N_c^2) \nonumber\\ 
+{2\pi a \over mV}\sum_{\vec{k},\vec{k}'}'(N_{\vec{k}}N_{\vec{k}'}
+M_{\vec{k}}M_{\vec{k}'})
\end{eqnarray}
where $M_{\vec{k}}=\delta_{\vec{k},\vec{0}}N_c+\sinh(2s_{\vec{k}})/2$ and 
where the prime in the sum remind us the effect of the operator 
$( \partial / \partial r)r$ in the pseudopotential to remove 
ultraviolet divergences \cite{Huang}. 
Minimizing the grand canonical Hamiltonian 
$\langle\hat H - \mu \hat N \rangle$ where $\mu$ is the 
chemical potential, we obtain the HFB equations \cite{Griffin,Siggia} which 
can be rewritten as the closed set of equations:
\begin{eqnarray}\label{s}
\sinh(2s_{\vec{k}})&=&
-{\Sigma_{12} \over 
\sqrt{({\vec{k}^2 /2m}-\mu + \Sigma_{11})^2-\Sigma_{12}^2}}
\\ \label{nc}
\mu&=&\Sigma_{11}+\Sigma_{12}-{8 \pi a \over m}n_c
\end{eqnarray}
where we have introduced the normal and anomalous self-energies 
respectively \cite{Hohenberg,Griffin,divergence}:
\begin{eqnarray}
\Sigma_{11}&=&{8 \pi a \over m}n
\\ \label{HP} 
\Sigma_{12}&=&\Sigma_{11}-\mu +
{8 \pi a \over V} \left(
\sum_{\vec{k}}{\sinh(2s_{\vec{k}})\over 2m}
+\sum_{\vec{k}\not= 0} {\Sigma_{12} \over \vec{k}^2} \right)
\end{eqnarray}
Here, $n=N/V$ is the particle density and $n_c=N_c/V$ is the coherent 
particle density.

The longwavelength excited states 
are given by 
$|\theta\rangle_{e,\vec{k}}=\hat D^{\dagger} \hat S^{\dagger}
e^{i\theta}\hat{c}^{\dagger}_{\vec{k}}|0\rangle$ from which we 
can estimate using  
(\ref{s}-\ref{HP}) the longwavelength excitation spectrum  
$_{e,\vec{k}}\langle\theta|\hat H | \theta \rangle_{e,\vec{k}}-E_g\sim
[({\vec{k}^2 /2m}-\mu + \Sigma_{11})^2-\Sigma_{12}^2]^{1/2}$.
In the particular HFB framework,
the Hugenholtz-Pines (HP) theorem states that $\mu = \Sigma_{11}-
\Sigma_{12}$ \cite{Hohenberg,Griffin} which is precisely the condition to have 
a gapless theory i.e. a theory in which there is no gap of energy between the
ground state and the longwavelength excited states.
In that case, 
the excitation spectrum 
corresponds to a phonon excitation
of momentum $\vec{k}$
with a velocity
$v=\sqrt{4 \pi a n_c/m^2}$.
An inspection of (\ref{HP}) shows that the HP theorem requires the term 
in parentheses of (\ref{HP}) to be 
either neglected as done in the Bogoliubov and Popov approximation schemes 
\cite{Popov} or either 
equal to zero.

The essential point of this paper is that the term in parentheses of (\ref{HP})
can be exactly zero if we macroscopically squeeze the condensed part state.

Indeed,  
we take for the parameter $s_{\vec{0}}$ a negative huge value in such a way 
that $n_s=N_s/V=(\cosh(2s_{\vec{0}})-1)/(2V)$ is a finite quantity  
different from zero in the thermodynamic limit \cite{Roepke}. Since 
by (\ref{s}) we have also the singular expression 
$\cosh^2(2s_{\vec{0}})=1/[1-\Sigma_{12}^2/(\Sigma_{11}-\mu)^2]$,
the existence of a 
macroscopic parameter $N_s$ respects automatically the HP theorem  
in the thermodynamic limit \cite{idealgas}. Thus, defining 
$\hat N_0=\hat c_{\vec{0}}^\dagger \hat c_{\vec{0}}$, 
the condensed fraction density $n_0=\langle \hat N_0 \rangle/V$ 
is not $n_c$ but the sum $n_0=n_c+n_s$   
and $\sinh(2s_{\vec{0}})/2=
-[n_s+1/(2V)]\Sigma_{12}/(\Sigma_{11}-\mu)\simeq -n_s$. The density 
$n_s$ is still an arbitrary 
parameter and must be ajusted to verify the HP 
relation in (\ref{HP}). Removing the zero momentum mode, 
we can replace the sum 
$\sum_{\vec{k} \not= 0}$
by the integral ${V \over (2 \pi)^3}\int d^3 \vec{k}$ in Eq.(\ref{HP}).
The integral can be easily carried out and gives, with the help 
of the HP relation and the Eq.(\ref{nc}), the result
$n_s=8(a^3n_c/\pi)^{1/2}n_c$. 

Similarly defining $\hat N_e=\hat N - \hat N_0$, the same replacement as 
above gives for the density of non-zero momentum particles  
$n_e=\langle\hat N_e\rangle /V= 8(a^3n_c/\pi)^{1/2}n_c/3$ 
which is three times 
smaller than $n_s$. Combining these results, we can solve (\ref{s}-\ref{HP}) 
to get closed relations between the chemical potential, 
the total particle density and the ground state energy: 
\begin{eqnarray}\label{mu}
\mu&=&{4\pi a n_c\over m}\left[1+{64 \over 3}({a^3 n_c \over \pi})^{1/2}\right]
\\ \label{n}
n&=&n_c\left[1+{32 \over 3}({a^3 n_c \over \pi})^{1/2}\right]
\\
{E_g\over V}&=&{2 \pi a n_c^2 \over m}
\left[1+{448 \over 15}({a^3 n_c \over \pi})^{1/2}+
{2048 \over 9}({a^3 n_c \over \pi})\right]
\\ \label{HF}
&\cong& {2 \pi a n^2 \over m} \left[1+
{128 \over 15} ({a^3 n \over \pi})^{1/2}+ {\cal O} ({a^3 n \over \pi}) \right]
\end{eqnarray}
These are related through the thermodynamical relation $d E_g /dN =\mu$.
Thus, up to the first order we recover the classical results derived 
in standard textbooks \cite{Huang,Fetter}.

The physical interpretation of 
the squeezing of the condensed part is as follows \cite{Walls}. 
Let us define the quadrature amplitudes 
$ \hat {\cal P}= (\hat c_{\vec{0}} + \hat c_{\vec{0}}^\dagger)/2$ 
and
$ \hat {\cal Q}= (\hat c_{\vec{0}}^\dagger - \hat c_{\vec{0}})/(2i)$ 
of the zero mode operators. 
The average value of the two operators are 
$\langle \hat {\cal P}\rangle=\cos \theta N_c^{1/2}$ and
$\langle \hat {\cal Q}\rangle=-\sin \theta N_c^{1/2}$ \cite{order}
while the quantum square fluctuations around their averages are much 
smaller in absolute value
$\langle \delta^2  \hat {\cal P} \rangle \cong
\sin^2 \theta N_s$ and 
$\langle \delta^2 \hat {\cal Q} \rangle \cong
\cos^2 \theta N_s$ ($\delta \hat A=\hat A - \langle \hat A \rangle$).
Representing these results on a phase space,
we see that 
the quantum fluctuations  
are the largest in the angular direction and microscopic 
in the radial one.

Considering the potential energy part  
$U(\hat c_{\vec{0}}^\dagger,\hat c_{\vec{0}})
=-\mu \hat c_{\vec{0}}^\dagger\hat c_{\vec{0}} + {2\pi a \over m V}
(\hat c_{\vec{0}}^\dagger\hat c_{\vec{0}})^2$ of the grand canonical 
Hamiltonian, 
the 
distribution of the quadrature amplitudes values prefers to spread out in 
a region where the potential energy is minimum \cite{Villain}.
Comparing with \cite{Lewenstein}, in the HFB approach,
$\hat {\cal P}$
is nearly a conserved quantity for $\theta=0$ 
since the fluctuations of $\hat {\cal P}$ are
microscopic and should be rather associated with {\it a collective motion with a
microscopic restoring force}. The commutation relations 
$[\hat {\cal Q}, \hat {\cal P}]=i/2$
imply that
the fluctuations of $\hat {\cal Q}$ are huge but finite.
                        
The only difficulty in this new approach is that the square quantum 
fluctuations of the condensed part population are not normal (i.e. linear 
in $N$ ) but 
enormous: $\langle \delta^2 \hat N_0 \rangle=
2N_s^2 \propto N^2$.
This is to be compared with the square quantum fluctuations 
of the excited states population 
which are normal \cite{Stringari}: 
\begin{equation}\label{dN}
\langle \delta^2 \hat N_e \rangle=
\sum_{\vec{k} \not=\vec{0}}
\sinh^2(2s_{\vec{k}})/2=2(\pi a^3n_c)^{1/2}N_c \propto N
\end{equation}
On the other hand, the particle number conservation imposes that these 
fluctuations should be equal since the population distributions are 
constrained by the relation $N_0+N_e=N$. 

This apparent paradox can be 
easily understood if one remembers that $|\theta\rangle_g$ is not 
an eigenstate of the total number operator $\hat N$ but has enormous  
fluctuations dominated by $\langle \delta^2 \hat N_0 \rangle$. 
The ansatz is not $U(1)$ 
invariant since the phase factor $\theta$ is fixed. Thus, in the 
HFB approach, we do not have one ground state solution but an infinite 
number of degenerate ones parametrized by $\theta \in 
[0,2\pi[$. A improved ground state solution is the superposition 
$|\phi\rangle_g=\int_0^{2\pi}{d\theta \over 2\pi} f(\theta)|\theta\rangle_g$ where 
the function $f(\theta)$ is determined by minimizing the energy 
in the $|\theta\rangle_g$ subspace.
In this way, we find the integral equation 
$\int_0^{2\pi}{d\theta \over 2\pi} 
\,_g \langle\theta'|(\hat H - E_0)f(\theta)|\theta\rangle_g=0$ with the 
eigenvalue $E_0$ which, using the 
properties $_g\langle\theta'|\theta\rangle_g=
 _{\ g}\!\!\langle0|\theta-\theta'\rangle_g$ and 
$_g\!\langle\theta'|\hat H|\theta\rangle_g=
 _{\ g}\!\!\langle0|\hat H|\theta-\theta'\rangle_g$,
becomes a convolution equation and can be solved by making an 
appropriate Fourier transform. The new solution 
is an eigenvector of $\hat N$ and reads:
\begin{equation}\label{phi}
|\phi_N\rangle_g={1 \over r_N} \int_0^{2\pi}{d\theta \over 2\pi}
e^{-iN\theta}|\theta\rangle_g
\end{equation} 
where $r_N$ is a normalisation factor.

Let us examine the property of 
this improved solution. 

First, it lowers the ground state energy. Indeed, due 
to the large fluctuations, the calculations show that the term 
${2 \pi a \over m V} \hat N_0^2$ contained in the Hamiltonian 
gives a larger contribution for $|\theta\rangle_g$ than for $|\phi_N\rangle_g$.
Consequently, the new ground state energy reads:
\begin{eqnarray}\label{E0}
{E_0 \over V}&=&{E_g \over V}- {4\pi a \over m}n_s^2
\\ \label{E01}&=&
{2 \pi a n_c^2 \over m}
\left[1+{448 \over 15}({a^3 n_c \over \pi})^{1/2}+
{896 \over 9}({a^3 n_c \over \pi})\right]
\end{eqnarray}    
We immediatlely check that this shift in energy does not affect 
the result (\ref{HF}). This demonstrates that, although in the HFB approach the 
symmetry is broken and the number fluctuations is enlarged, ultimately 
the ground state should conserve the particle number. 
Had we tried the more elaborated expression (\ref{phi}) as the ansatz,
we would have obtained, after a more technical calculation, exactly the 
same result (\ref{E0}). Similarly, we can build the number conserving 
excited states $|\phi_N\rangle_{e,\vec{k}} \sim \int_0^{2\pi}{d\theta \over 2\pi}
e^{-iN\theta}|\theta\rangle_{e,\vec{k}}$ which are orthogonal to (\ref{phi}) 
and check that the excitation spectrum is unchanged.

Second, it is interesting to see how the condensed part is 
populated. This will illustrate what are the technicalities involved to 
handle with (\ref{phi}).
To this end, we notice that $|\theta\rangle_g$ is a generating state of 
$|\phi_N\rangle_g$ which further can be decomposed into a product of 
generating states, one for the condensed part state
$|N_0\rangle=(N_0!)^{-1/2}\hat c_0^{\dagger N_0} |0\rangle$, the other for the  
non-zero momentum pairs 
state $|N_e\rangle$:
\begin{eqnarray}\label{genst}
|\theta\rangle_g &=&\sum_{N=0}^\infty e^{iN \theta} r_N |\phi_N\rangle_g
\nonumber \\
&=& 
\sum_{N_0=0}^\infty e^{iN_0 \theta} p_{N_0} |N_0\rangle \otimes
\sum_{N_e=0}^\infty e^{iN_e \theta} q_{N_e} |N_e\rangle
\end{eqnarray}
The amplitude of the coefficients $r_N$, $p_{N_0}$ and $q_{N_e}$ can 
be determined through their respective generating functions:
\begin{eqnarray}
_g\langle0|\theta\rangle_g &=&\sum_{N=0}^\infty e^{iN \theta}|r_N|^2
=G_0(\theta) G_e(\theta)
\\
G_0(\theta)&=&\sum_{N_0=0}^\infty e^{iN_0 \theta} |p_{N_0}|^2
=\left[{1-\tanh^2 s_{\vec{0}} \over 1-
e^{2i\theta}\tanh^2 s_{\vec{0}}}\right]^{1/2}
\nonumber \\ 
&\times&\exp \left[{1-\tanh s_{\vec{0}} \over 
1- e^{i\theta}\tanh s_{\vec{0}}}(e^{i\theta}-1)N_c \right]
\\
G_e(\theta)&=&\sum_{N_e=0}^\infty e^{iN_e \theta} |q_{N_e}|^2 
\nonumber\\& =&
\prod_{\vec{k}\not= \vec{0}}
\left[{1-\tanh^2s_{\vec{k}} \over 1-e^{2i\theta}\tanh^2 s_{\vec{k}}}
\right]^{1/2}
\end{eqnarray}
An exact explicit expression for $p_{N_0}$ can be calculated 
\cite{Walls} while an approximate one for $q_{N_e}$ is 
found using the saddle point method \cite{Navez}.
An inspection of (\ref{genst}) shows that the probability distribution 
of having $N_0$ condensed particles out of $N$ in the state 
$|\phi_N\rangle_g$ is given by 
$P(N_0)=|p_{N_0} q_{N-N_0}/ r_N|^2$.
More explicitely,
\begin{eqnarray}
P(N_0)&\cong&
{{\cal N}\over N_0!} \left(-{\tanh s_{\vec{0}} \over 2} \right)^{N_0}
H^2_{N_0}({(1-\tanh s_{\vec{0}})N_c^{1/2}
\over (-2 \tanh s_{\vec{0}})^{1/2}})
\nonumber\\
&\times&\exp[ -
{(N_0-\langle \hat N_0 \rangle)^2 \over 2 \langle \delta^2 \hat N_e \rangle}]
\end{eqnarray}
where $H_n(x)$ are the Hermite polynomials and ${\cal N}$ is 
a normalisation factor and $\langle \delta^2 \hat N_e \rangle$ 
is given in Eq.(\ref{dN}).
The population distribution is not a poissonian one but a 
squeezed state one \cite{Walls}
whose the intensity is modulated by 
a gaussian envelope centered in $\langle \hat N_0 \rangle$.
The particle number fluctuations of the condensed part is given by 
$\langle \delta^2 \hat N_0 \rangle_N= 
\,_g\langle \phi_N | \delta^2 \hat N_0 | \phi_N \rangle_g=
\langle \delta^2 \hat N_e \rangle_N$.
Again using the saddle point method, we calculate \cite{Navez}:
\begin{eqnarray}
\langle \delta^2 \hat N_0 \rangle_N &=&
{ \partial^2 \over \partial \alpha^2}
\ln [\int_0^{2\pi} \!\!\!\!\!\! d\theta\, e^{-iN\theta}
G_0(\theta)G_e(\theta+i\alpha)]_{\alpha=0}
\\
&\cong& \langle \delta^2 \hat N_e \rangle=
2(\pi a^3n_c)^{1/2}N_c
\end{eqnarray}
showing that the fluctuations are indeed normal. This result 
is identical to that derived previously by Giorgini et al. 
\cite{Stringari} by considering the condensed part as a 
particle reservoir.

In conclusion, we have constructed a ground state which is 
stationary in the HFB approach. The key ingredient is the 
macroscopic squeezing of the condensed part.
This squeezing implies prohibitively large 
particle number fluctuations which are suppressed if we
build the appropriate $U(1)$ invariant number 
conserving superposition of the squeezed states.  

An important issue 
is to know how the macroscopic squeezing is modified 
at non zero temperature and how it changes the critical temperature 
\cite{Gruter} as well as other quantities of experimental interest 
\cite{Clark}. 
This issue can be addressed by means of a variational 
approach \cite{Siggia,Stoof} in the same way as for 
fermions in the theory of superconductivity \cite{BCS}. 
Other topics include: the investigation of the validity of
the non-perturbative mean field result (\ref{E0}) according to the 
coupling parameter $na^3$ \cite{Gruter} and 
an analyse of
the quantum phase diffusion,
in particular, an improved calculation of the two-time correlation function
$\langle \hat \psi^\dagger (\vec{r},t) \hat \psi(\vec{r},0) \rangle_N$
\cite{Lewenstein,Wilkens}. These further developments would allow to find a test for the
experimental evidence of the macroscopic 
squeezing.

{\bf ACKNOWLEDGMENTS} P.N. has been supported by the TMR Network 
"Coherent Matter Wave Interactions", 
contract ERBFMRX CT96-0002 and by the Republic of 
Greece State Scholarship Foundation, contract No 384.
P.N. wants to acknowledge constructive criticism by Professor M. Wilkens and 
to thank Professor A. Theophilou and Dr. C. Henkel for 
enlightening discussions.

\end{document}